\documentclass[12pt]{iopart}

\usepackage{iopams}
\usepackage{graphicx}
\usepackage{bm}

\newcommand{\Z}{\mathbb{Z}}
\newcommand{\N}{\mathbb{N}}
\newcommand{\beq}{\begin{equation}}
\newcommand{\eeq}{\end{equation}}

\newcommand{\n}{\mathbf{n}}
\newcommand{\kk}{\mathbf{k}}
\newcommand{\x}{\mathbf{x}}
\newcommand{\eps}{\varepsilon}
\newcommand{\mafrac}{\sqrt{\frac{8 \pi}{3}}}

\begin{document}
\title[Dynamics of a lattice Universe]{Dynamics of a lattice Universe.\\
{\it The dust approximation in cosmology.}}

\author{Jean-Philippe Bruneton$^1$ and Julien Larena$^2$}
\address{$^1$ Namur Center for Complex systems (naXys), University of Namur, Belgium}
\address{$^2$ Department of Mathematics, Rhodes University, Grahamstown 6140, South Africa.}
\ead{jpbr@math.fundp.ac.be; j.larena@ru.ac.za}

\begin{abstract}
We find a solution to Einstein field equations for a regular toroidal lattice of size $L$ with equal masses $M$ at the centre of each cell; this solution is exact at order $M/L$. Such a solution is convenient to study the dynamics of an assembly of galaxy-like objects. We find that the solution is expanding (or contracting) in exactly the same way as the solution of a Friedman-Lema\^itre-Robertson-Walker Universe with dust having the same average density as our model. This points towards the absence of backreaction in a Universe filled with an infinite number of objects, and this validates the fluid approximation, as far as dynamics is concerned, and at the level of approximation considered in this work. 
\end{abstract}



\section{Introduction}
In modern cosmology, the large scale structure of the Universe is usually described by a Friedmann-Lema\^itre-Robertson-Walker (FLRW) geometry, associated, via the Einstein field equations, with a perfectly homogeneous and isotropic distribution of matter. The FLRW geometry supplemented by perturbation theory proves to be an excellent model for the Universe on large scales, considering the wide range of independent observations that it can fit. Nevertheless, this ability to fit most, if not all, the observations available today comes at a cost: the unavoidable introduction of dark matter and dark energy: two elusive energy components that actually dominate the dynamics of the late time Universe on all scales. The introduction, by hand, of these new, exotic, components thus raises the concern that something may be wrong in the more fundamental way we describe the late-time Universe, i.e. it inevitably leads one to wonder if our theory of gravitation and/or our assumptions on the geometry of the Universe might be oversimplified. The first possibility advocates for a modification of the laws of gravity on large scale (see, for example \cite{Clifton:2011jh} for a recent review). Weaker assumptions on the geometry of the Universe have also been explored, mainly in the context of Lema\^itre-Tolman-Bondi models; see, for example \cite{Celerier:1999hp,Alnes:2005rw,Enqvist:2007vb,GarciaBellido:2008nz,February:2009pv,Clarkson:2010ej}, but also \cite{Moss:2010jx,Moss:2011ze} for criticisms of this possibility.\\
\\
Of course, the actual Universe, at least in its later stage of evolution, is not perfectly homogeneous and isotropic, because of the gravitational instability that causes structure to form out of tiny initial fluctuations in the matter distribution, forming bound objects such as galaxies and clusters of galaxies. Therefore, the FLRW geometry is usually constructed by representing, in the field equations, the complicated, inhomogeneous energy-momentum content by a smoothed energy-momentum tensor involving averaged, non local, energy densities and pressures. This is the so-called fluid approximation: on large scales, the energy content of the Universe is well modelled by one or several homogeneous and isotropic perfect fluids. Despite the fact that it is widely used, it is not very often stated explicitly, let alone constructed (see \cite{Wiltshire:2011vy} for a recent discussion). Moreover, it is now well-known that averaging cosmological quantities leads to the introduction of backreaction terms \cite{Ell84,Buchert:1999er,Buchert:1999mc,Buchert:2001sa} that could modify the kinematics of the model on large scales, even though the amplitude of such corrections is not presently determined and is still a matter of debate (see \cite{Clarkson:2011zq} for a critical discussion on these issues).\\
\\
The purpose of this paper is not directly to discuss the backreaction issue. Rather, we would like to propose a solution of Einstein field equations corresponding to a set of point masses representing objects such as galaxies, in gravitational interaction with each other in an otherwise empty spacetime. This will allow us to discuss the kinematics of the collection of point masses and compare it to the kinematics of an analogue FLRW model with a fluid corresponding to the one emerging from the distribution of masses when it is smoothed.  This route has recently been explored by Clifton and Ferreira in a series of papers \cite{Clifton:2009jw,Clifton:2009bp} based on the Lindquist-Wheeler tessellation of spacetime \cite{LindWheel}. The Lindquist-Wheeler approximation consists in pasting together a large number of Schwarzschild cells representing the metric around point masses. Then, the Israel junction conditions imply that the boundary on which the cells are glued expands following the FLRW kinematics. Unfortunately, it is impossible to glue these cells together exactly: the Israel junction conditions are only satisfied on a subset of the boundaries, where they intersect, but the boundaries cannot be exactly matched, leading to an overlap region and a 'no-man's land' \cite{Clifton:2009jw}. Also, spherical symmetry in each cell is only approximate. Therefore, despite its elegance, this method relies on approximations that are difficult to control, and one would hope for symmetric situations in which an exact solution could be found. This was done in \cite{Uzan:2010nw}, where two antipodal masses were put on a 3-sphere. It was shown that an exact solution to this problem could only be found by introducing a cosmological constant. Recently, and independently of the current work, \cite{Clifton:2012qh} have studied a more realistic solution in the context of cosmology, namely, a finite number of masses arranged in a regular lattice on a hypersurface having the topology of a 3-sphere. Nevertheless, they only studied the kinematical backreaction arising at a given time (the time of maximum expansion) by solving the constraints equations, leaving the full dynamics for further studies. A similar study, but for the flat topology was also conducted numerically in \cite{Yoo:2012jz}.\\
\\
In this paper, we will show that, by considering an infinite cubic lattice of size $L$, with equal masses $M$ at the centre of each cell, one can find a solution to the Einstein field equations that is exact at second order in an expansion in powers of $\sqrt{M/L}$. Such an approximation scheme improves on the Lindquist-Wheeler approximation, since there is no need for junction conditions (and hence, no 'no-man's land'), and because it retains a certain degree of anisotropies around each mass. The main result of the paper is thus the expression for the metric, Eq.~(\ref{mainresultmetric}). A similar type of solution was studied in \cite{Korotkin:1994dw}, with periodicity along one axis. \cite{Clifton:2010fr} also investigated such lattices, by concentrating on small deformations of the geometry of each cell in powers of $v/c$, the velocity of the objects described by the lattice. Our approach is very similar to this work, but instead of solving the perturbed junction conditions between cells, we solve directly for the full metric of spacetime perturbatively.\\
\\
The choice of $\sqrt{M/L}$ as a small parameter will be motivated by the fact that a regular Taylor expansion of the metric in powers of $M/L$ cannot satisfy the field equations in a toroidal symmetry. Rather, one has to rely on an expansion in terms of $\sqrt{M/L}$, because in this case, the source terms in the equations, that is of order $M/L$, can be reabsorbed by a correct re-definition of the integration constant at order $\sqrt{M/L}$, leading to a well-behaved solution. It will also be shown that, by the same process, the scalar solution at order $M/L$ is fully determined by requiring that the equations be satisfied at order $\left(M/L\right)^{3/2}$. In Sect. \ref{sec2}, we present the solution obtained when the distribution of masses is modelled by a Dirac comb. We discuss the subtleties associated with the series expansion and present the method used to obtain the solution in details. We will see that the Fourier expansion of the solution leads to diverging terms in the metric in the form of non convergent series exhibiting ultraviolet divergencies. Therefore, there is no solution to the problem for a singular distribution of masses, in agreement with the results of \cite{Korotkin:1994dw}. In Sect. \ref{sec3}, we address this non-convergence problem by smearing the distribution of mass. This amounts to replacing the Dirac comb by a sum of approximations of the Dirac $\delta$ function via peaked Gaussian functions. This will introduce a natural UV cut-off in the diverging parts of the metric, and this cut-off will be set properly to obtain a solution that describes accurately the spacetime exterior to the masses. In Sect. \ref{sec4}, we discuss the kinematical characteristics of the solution via its the effective scale factor, and we compare this one to the analogue FLRW model with dust, where the energy density of the dust fluid is identical to the one derived from the distribution of point-like mass in our lattice. We find that the fluid approximation is valid for the special solution presented in this paper, since the scale factors derived from the exact solution and from the analogue FLRW model match at order $(M/L)^{3/2}$. This is in agreement with the results of \cite{Clifton:2012qh}, since they found that the kinematical backreaction in their finite model decreases as the number of masses increases: here we have an infinite number of masses, so that at the limit, we would indeed expect the backreaction effect to exactly cancel. Nevertheless, this comparison should be made with caution, since we only have a perturbative solution at order $(M/L)^{3/2}$; non-linear effects at higher orders could generate backreaction through mode couplings.  Finally, Sect. \ref{sec5} serves as a concluding section and introduces some future works.\\
\\
We use the sign convention $(-,+,+,+)$ for the spacetime metric. Latin indices are used for space components, and boldface symbols denote three vectors as well as triplets of integers. We also use natural units $G=c=1$ throughout the paper, except stated otherwise.

\section{A lattice Universe made of point masses}
\label{sec2}
\subsection{Framework}
Physically, we are trying to describe the collective behaviour of typical astrophysical objects such as galaxies. This is clearly an impossible task if we do not introduce further assumptions on the symmetries of the problem. The idea is to obtain a model that is genuinely inhomogeneous, but exhibits an averaged homogeneity when considered on large scales. A regular lattice of identical masses certainly possesses these characteristics, and we will consider a cubic lattice of size $L$, with identical masses $M$ at the centre of each cell of the lattice. Therefore, space clearly has a toroidal symmetry. The periodicity of the distribution of sources will be extremely valuable when trying to solve the field equations, since it will impose a periodic solution, and allow the use of Fourier series.\\
\\
Nevertheless, even formulated as such, it is extremely difficult to solve the full problem, because of the non-linearity of the field equations. But we can do one further assumption based on the physics we are trying to understand. If the masses on the lattice are to represent typical galaxies, we can choose, as our typical parameters
$$
M\sim 10^{11}M_{\odot} \mbox{ and } L\sim 1\mbox{ Mpc},
$$ 
where $M_{\odot}\sim 10^{30}$ kg is the Solar mass, and $L$ is of order of the intergalactic distances. With these numbers, the natural parameter of the lattice is $R_{S}/L\sim 10^{-8}\ll 1$, where $R_{S}=2M$ is the Schwarzschild radius of the masses. Therefore, we can look for a solution expanded into powers of $M/L$; this will lead to linearised field equations that can be solved exactly, and we expect that the first few terms in the solution will represent the spacetime exterior to the masses with high accuracy. 

\subsection{Definition of the perturbative expansion}
In the following we will consider a source term given by a three dimensional Dirac comb in comoving coordinates $T_{00} \propto M \sum_{\n \in \Z^3} \delta^{(3)}(\x-L\n )$, where the masses are located at $\x_{\n}=L\mathbf{n}$. Such a periodic distribution of masses can be written as a Fourier series with 
\beq
\label{Source}
M \sum_{\n \in \Z^3} \delta^{(3)}(\x-L\n )=\frac{M}{L^3}\sum_{\n \in \Z^3} e^{i \frac{2 \pi \mathbf{n}.\x}{L}},
\eeq
showing that the zero mode of the source term is a constant nonzero (comoving) density $\rho=M/L^3$. We could look for a periodic solution for the metric of the standard form: $g_{\mu\nu}=\eta_{\mu\nu}+h_{\mu\nu}$, where $h_{\mu\nu}$ would be of order $M/L$. However in this case,  one finds, Fourier expanding $h_{\mu\nu}$, that $G_{00}^{\textrm{\small{zero mode}}} =\mathcal{O}((M/L)^2)$  showing that no periodic metric of the form considered above can solve Einstein equations for such a matter source.
\\
\\
This is reminiscent of an analogue situation for the Coulomb potential in infinite lattice of charges, i.e. crystals. In crystals the zero mode of the periodic Coulomb potential cannot be solved in general, and one simply drops this term by noticing that the average density of charges vanishes, provided the crystal is electrically neutral \cite{Crandall, Crandall2}, see also \cite{Marcos:2006cn}. In our case however, the global scalar charge does not vanish and the zero mode of Einstein equations, in a periodic solution, must be absorbed through an homogeneous expansion of space. Comparing to a flat and homogeneous, dust-dominated universe, we see that Hubble constant relates to the density of matter via $H^2(t)=8 \pi \rho/3$, showing that the corresponding FLRW metric contains a $\sqrt{\rho} \sim \sqrt{M/L^3}$ term.
\\
\\
As a consequence, we need instead to consider a perturbative expansion for the metric of the following form
\beq
\label{metricansatz1}
g_{\mu\nu}=\eta_{\mu\nu} + \sqrt{\frac{M}{L}} h^{(1)}_{\mu\nu}+\frac{M}{L} h^{(2)}_{\mu\nu}+\frac{M^{3/2}}{L^{3/2}} h^{(3)}_{\mu\nu}+\mathcal{O}\left(\frac{M^2}{L^2}\right),
\eeq
where the first order term will absorb the zero mode of the source term through an homogeneous expansion of space (i.e. we will select the solution for which $h^{(1)}_{\mu\nu}$ depends solely on time), whereas the anisotropic gravitational field created by the masses will be described in the $h^{(2)}_{\mu\nu}$ part. This makes clear from the beginning that a solution up to the order $3/2$ in $M$ is attainable, mixing the global expansion with the local anisotropic gravitational field. However the order $M^2$ solution is out of reach since Einstein equations at this order involve non linearities, and in particular product of Fourier sums and non-trivial mixing of the Fourier modes.  
With the ansatz Eq.~(\ref{metricansatz1}), we are now in position to derive both the expressions for the matter source and linearised field equations.

\subsection{Matter source}
The stress-energy tensor corresponding to our lattice universe reads
\beq
\label{Tab}
T^{\mu\nu}(t,\x)=  \frac{M}{\sqrt{-g(t,\x)}} \sum_{\n \in \Z^3} \frac{u^{\mu}_{\n}u^{\nu}_{\n}}{u^0_{\n}}\delta^{(3)}(\x- L\n ).
\eeq
Comoving coordinates require  $u^i_{\n}=0$ and therefore $u^{\mu}_{\n}=\delta^{\mu}_{0}/\sqrt{-g_{00}(\x_{\n})}$. Energy conservation imposes $\nabla_{\mu} T^{\mu\nu}=0$, and is automatically solved to any order by choosing the synchronous comoving gauge $g_{00}=-1$ and $g_{0i}=0$ in which we will work for now on. Then one shows that the only non-vanishing component of $T_{\mu\nu}$ to order $M^{3/2}$ reads
\beq
\label{Tab2}
T_{00}= M\left( 1-\frac{1}{2}\sqrt{\frac{M}{L}} \sum_{i} h_{ii}^{(1)}\right)\sum_{\n \in \Z^3}\delta^{(3)}(\x- L\n ),
\eeq
where we used Eq.~(\ref{Tab}), the gauge conditions, the fact that $T^{\mu\nu}$ is $\mathcal{O}(M)$ to the leading order, and $g_{\mu\nu}=\eta_{\mu\nu} + \sqrt{\frac{M}{L}} h^{(1)}_{\mu\nu} +\mathcal{O}(M)$. Fourier expanding the Dirac Comb, we finally get
\beq
T_{00} = \frac{M}{L^3} \left( 1-\frac{1}{2}\sqrt{\frac{M}{L}} \sum_{i} h_{ii}^{(1)}\right) \sum_{\n \in \Z^3} e^{i \frac{2 \pi \mathbf{n}.\x}{L}} +\mathcal{O}\left(\frac{M^2}{L^2}\right).
\eeq
In the following we shall also work with the wave vector $\kk= 2 \pi \n/L$ and write $\sum_{\n \in \Z^3} e^{i \frac{2 \pi \mathbf{n}.\x}{L}} \equiv \sum_{\kk} e^{i \kk.\x}$.

\subsection{Metric, field equations and solution}
In the synchronous gauge comoving with the masses, we split the line element w.r.t. scalar, vector, and tensorial parts
\begin{equation}
\label{MetricSVT}
ds^{2}=-dt^{2}+\left[(1+2C)\delta_{ij}+2\partial_{ij}^{2}E+2\partial_{(i}\bar{E}_{j)}+\bar{E}_{ij}\right]dx^{i}dx^{j},
\end{equation}
where, according to Eq.~(\ref{metricansatz1}), we write $C= \sqrt{M/L} C^{(1)}+ (M/L) C^{(2)} + (M/L)^{3/2} C^{(3)}$ and similarly for the other quantities. Then divergence free and tracelessness split accordingly into $\partial_{i}\bar{E}^{i (p)}=0$, $\partial_{i}\bar{E}^{(p) ij}=0$ and $\bar{E}^{(p) i}_{i}=0$ for $p=(1,2,3)$.
Einstein equations are decomposed into their scalar, vector and tensor parts. 
\subsubsection{Order  $\sqrt{M/L}$ field equations}
We get, for all $(i,j)$, and where a prime denotes a derivative with respect to $t$
\begin{itemize}
\item Scalar part
\begin{eqnarray}
0&=&\Delta C^{(1)}\\
0&=&\partial_{i}C^{(1)\prime}\\
0&=&\partial^{2}_{ij}\left(E^{(1)\prime \prime} - C^{(1)}\right) -\Delta \left(E^{(1)\prime \prime} - C^{(1)}\right)\delta_{ij} - 2 C^{(1)\prime \prime} \delta_{ij},
\end{eqnarray}
\item Vector part
\begin{eqnarray}
\label{VectorEq1}
0&=&\Delta\bar{E}^{(1)\prime}_{i}\\
\label{VectorEq2}
0&=&\partial_{i}\bar{E}^{(1)\prime \prime}_{j}+\partial_{j}\bar{E}^{(1)\prime \prime}_{i},
\end{eqnarray}
\item Tensor part
\begin{equation}
\label{TensorEq}
\bar{E}^{(1)\prime \prime}_{ij}-\Delta\bar{E}^{(1)}_{ij}=0.
\end{equation}
\end{itemize}
We look for periodic functions and write, e.g.  $C^{(1)}(t,\x)=C^{(1)}(t)+\sum_{\kk \neq 0} C^{(1)}_{\kk}(t) e^{i \kk \x}$. The system of equations is then solved mode-by-mode and one easily gets the most general solution
\begin{eqnarray}
C^{(1)}&=& \frac{A}{L} t + B \nonumber \\
E^{(1)}&=&L^2 E^{(1)}(t) + L K(\x) t + L^2 J(\x) \nonumber\\
\bar{E}^{(1)}_{i}&=& \bar{E}^{(1)}_{i}(t)\nonumber\\
\label{GravWaves12}
\bar{E}^{(1)}_{ij}&=& \alpha_{ij} t +\beta_{ij} + \sum_{\kk \neq 0} A_{ij}^{\kk} e^{i (\kk \x - |\kk|t)},\nonumber
\end{eqnarray}
where $K$ and $J$ are any dimensionless  zero mode free periodic functions. $E^{(1)}(t)$ and $\bar{E}^{(1)}_{i}(t)$ are not determined but do not appear in the expression for the metric, and can thus be taken to zero without loss of generality. Physically, we see on the above equations that the vector modes are not sourced and do not propagate, explaining why $\bar{E}^{(1)}_{i}=0$.  On the other hand the tensor modes propagate but are not sourced, so that we recover the standard gravitational waves. \\
\\
Derivatives of the functions $K$ and $J$ appear in the metric. These terms have to be understood as being the most general source terms compatible with the symmetries and which do not contribute to Einstein field equations to this order. They are ``spurious" source terms with no physical counterparts. Indeed it is easily shown that they act as source terms for order $M$ equations, and enter these equations in a non linear way. It is thus physically meaningful to set them to zero (and anyway necessary to be able to proceed further in the perturbative expansion). The same remark applies to the gravitational waves found above. First they have no clear physical origin, as they are of order $\sqrt{M}$, and second, if not dismissed, they would source the tensorial field equations to order $M$ in a non trivial way. 
\\
\\
We thus restrict ourselves to the simplest solution, namely $C^{(1)}= A/L t + B$, $E^{(1)}=0$, $\bar{E}^{(1)}_{i}=0$ and $\bar{E}^{(1)}_{ij}=0$ and proceed to the next order. As discussed previously, the solution we picked up only depends on $t$ and essentially absorbs the zero mode of the source term at order $M$, via the constant $A$. This is what we show now.

\subsubsection{Order  $M/L$ field equations} 

Let us start with the vector and tensorial equations. Using the previous $\sqrt{M}$ solution, they are found to be the same as Eqs.~(\ref{VectorEq1}, \ref{VectorEq2}, \ref{TensorEq}) up to the fact that it now applies to second order quantities  $\bar{E}^{(2)}_i$ and $\bar{E}^{(2)}_{ij}$. Thus, the vector part vanishes, while we choose to dismiss again the gravitational waves\footnote{One could keep these gravitational waves, as they will only source the order $M^2$ field equations, which go beyond the present solution. However we are more interested in the gravitational field created by the masses and the subsequent expansion of the lattice rather than by its possible content in gravitational waves.}. We therefore focus on the scalar equations, which are
\begin{eqnarray}
\label{EqS1}
3 \frac{A^2}{L^2}- 2 \Delta C^{(2)} = \frac{8 \pi }{L^2} \sum_{\kk} e^{i \kk \x} & & \\
\label{EqS2}
\partial_{i}C^{(2)\prime}=0 & & \\
\label{EqS3}
\partial^{2}_{ij}\left(E^{(2)\prime \prime} - C^{(2)}\right) -\Delta \left(E^{(2)\prime \prime} - C^{(2)}\right)\delta_{ij}  &=&\left(2 C^{(2)\prime \prime}-\frac{A^2}{L^2} \right) \delta_{ij}
\end{eqnarray}
We see how the order $\sqrt{M/L}$ solution for $C$ ($C^{(1)}=A t /L +B$) now enter the order $M/L$ equations, and can absorb the zero mode of the source term (to this order). Zero and non zero modes of Eq.~(\ref{EqS1}) indeed give, respectively
\begin{eqnarray}
\label{SolA}
A= \eps \mafrac & &\\
\label{SolC2}
C^{(2)}(t,\x)= C^{(2)}(t)+ f(\x), & &
\end{eqnarray}
where $\eps =\pm 1$ and where the function $f$, central to our solution, is given by
\beq
\label{fdef1}
f(\x)= \frac{1}{\pi}\sum_{\n \in \Z^3_{*}} \frac{1}{|\n|^2} e^{i \frac{2 \pi \n. \x}{L} },
\eeq
and will be discussed below. The star in the notation $\Z^3_{*}$ means that the null triplet $(0,0,0)$ is dismissed. Equivalently we have
\beq
\label{fdef2}
f(\x)=\frac{8}{\pi}\sum_{(n,p,q)\in\N^{3}_{*}}\frac{\cos\left(\frac{2\pi}{L}nx\right)\cos\left(\frac{2\pi}{L}py\right)\cos\left(\frac{2\pi}{L}qz\right)}{n^{2}+p^{2}+q^{2}}.
\eeq
Eq.~(\ref{EqS2}) is then trivially satisfied, while Eq.~(\ref{EqS3}) gives both
\begin{eqnarray}
\label{SolC2t}
C^{(2)}(t)&=&\frac{2 \pi t^2 }{3 L^2} +\alpha \frac{t}{L} +\beta \\
\label{SolE2}
E^{(2)}(t, \x)&=& L^2 E^{(2)}(t) + \frac{t^2}{2} f(\x) + L U(\x) t + L^2 V(\x) ,
\end{eqnarray}
where again $E^{(2)}(t)$ is left unspecified but drops from the metric, whereas now $U$ and $V$ are undetermined arbitrary (periodic, zero mode free) functions. 

\subsubsection{Order  $(M/L)^{3/2}$ field equations} 
We will not detail here the equations of motion as they are more complicated. One can easily derive them using any tensorial package, and solve them in a similar fashion as was done for the previous orders. This is straightforward although a bit tedious. Focusing on the scalar part, we have been able to derive the most general (scalar) solution which is displayed in the Appendix, see Eq.~(\ref{EqFullMetric}). It is the ``most general solution'' provided that one sets $K$ and $J$ found in the $\sqrt{M}$ solution to zero, and dismissing the gravitational waves. In the course of solving order $3/2$ equations, it turns out that, in analogy with the transition between the solution at order $\sqrt{M/L}$ and the solution at order $M/L$, during which the constant $A$ acquired a value in order for the equations at order $M/L$ to be satisfied, the constant $\alpha$ in Eq.~(\ref{SolC2t}) has to take a specific value
$$
\alpha=\eps B\sqrt{\frac{2\pi}{3}}. 
$$
The integration constants $B$, $\beta$, $\Upsilon$, and $\varpi$ found in Eq.~(\ref{EqFullMetric}), if not dismissed, would enter the effective scale factor (see Sect. \ref{sec4}) at post-Newtonian order. These terms have thus no Friedmannian counterparts and it is natural to set them to zero. The arbitrary functions $U,V,R$ and $S$ (see Eq.~(\ref{EqFullMetric})) affect the local dynamics of spacetime, but again, are essentially source terms with no physical counterparts and would be constrained by the higher order field equations. We thus let them to zero and obtain the simplest $(M/L)^{3/2}$ solution for the lattice universe (putting back $G$ and $c$): $g_{00}=-1$, $g_{0i}=0$, and 
\begin{eqnarray}
\label{mainresultmetric}
g_{ij}= \delta_{ij} \left[1+ 2\eps\sqrt{\frac{GM}{Lc^2}}  \mafrac \frac{c t}{L} +\frac{2GM}{L c^2}\left(f(\mathbf{x})+ \frac{2 \pi c^2 t^2}{3 L^2} \right) \right.
\nonumber\\
\left. +  2 \left(\frac{GM}{Lc^2}\right)^{3/2}\left( 2 \epsilon  \frac{c t}{L} \mafrac  f(\mathbf{x})  - \frac{2 \pi \epsilon}{9}\mafrac \frac{c^3 t^3}{L^3}\right) \right] \nonumber\\
+\frac{G M}{L c^2} c^2 t^2\partial_{ij}^2 f(\mathbf{x}) 
+ \left(\frac{GM}{Lc^2}\right)^{3/2} \eps\sqrt{\frac{8 \pi}{3}} \frac{ c^3 t^3}{3 L} \partial_{ij}f(\x),
\end{eqnarray}
where the function $f$ was given in Eq.~(\ref{fdef2}) above, and we remind that $\eps=\pm 1$. We will show in Sect. \ref{sec4} that $\eps$ relates to the sign of the effective Hubble constant: $\eps=1$ corresponds to an initially expanding universe that decelerates under the effect of gravity, while $\eps=-1$ is an initially contracting Universe that accelerates its contraction.  This metric solves Einstein equations for the matter source given in Eq.~(\ref{Tab2}) up to order $(M/L)^{3/2}$. \\
\\
Clearly the solution is valid only in the limit where $g_{\mu\nu}-\eta_{\mu\nu} \ll 1$. Under the identification of the effective Hubble constant $H = \sqrt{8 \pi/3} \sqrt{M/L^3}$ (see Sect. \ref{sec4}), the time-dependent terms in the above metric show that our solution is valid locally in time $\Delta t \ll H^{-1} \propto L \sqrt{L/M}$, while space-dependent terms are more complicated to analyse because of the non trivial form of the function $f$. Still, based on dimensional analysis at least, we also expect our solution to be valid only over a similar range $\Delta x \ll H^{-1} \propto  L \sqrt{L/M}$. Note that, for the numbers quoted previously, corresponding to a lattice of galaxy-like objects, the time span valid for the solution is of order 1 Gyr, which is already a good cosmological time, allowing the use of this solution for cosmological predictions.\\
\\
The function $f$ and its derivatives are divergent, and so is the metric. However next section will show how to regularize it by introducing a physically motivated cut-off. It is enough, here, to note that the divergence in $f$ comes from large values of the Fourier modes and is therefore a UV divergence linked to the point-like nature of the sources. Therefore, if the sources are extended but still very small compared to $L$, we expect to obtain a similar solution. This will indeed be proven in the next section.

\section{A Lattice universe with peaked Gaussians}
\label{sec3}

The function $f(\x)$ and its derivatives are ill-defined, because the series expansion that defines $f(\x)$ does not converge. However, this problem comes from a UV divergence of the expansion, and can be easily dealt with. One should remember that one can approximate the Dirac $\delta$ function by
\begin{equation}
\delta(x-nL)\sim \frac{1}{\eta\sqrt{\pi}}e^{-\frac{(x-nL)^{2}}{\eta^{2}}},
\end{equation}
for $\eta$ arbitrarily small, and where $\eta$ has a dimension of a length. This $\eta$ will be related to a UV cut-off applied to Fourier modes. Then the (total) regularized source term reads $S(x,y,z)=S(x)S(y)S(z)$ with
\begin{equation}
S(x)=\sum_{n\in\Z}\frac{1}{\eta\sqrt{\pi}}e^{-\frac{(x-nL)^{2}}{\eta^{2}}}.
\end{equation}
Such a source term turns out to be the convolution of a Dirac Comb with a normal distribution of variance $\eta^2$. By virtue of the convolution theorem its Fourier transform is thus given by the product of the Fourier transforms of the Dirac Comb and the normal distribution. One then finds that the total source term can equivalently be written as
\begin{equation}
\label{SourceFourier}
S(x)S(y)S(z)=\frac{1}{L^{3}}\sum_{\mathbf{n}\in\Z^3}e^{\frac{2\pi}{L}i\mathbf{n}.\x-\frac{\pi^{2}|\mathbf{n}|^{2}\eta^{2}}{L^{2}}}.
\end{equation}
We also note that the source term is properly normalized. Indeed, taking advantage of Eq.~ (\ref{SourceFourier}), one can write
\beq
S(x)=1+ 2\sum_{n=1}^{\infty} e^{-\frac{n^2 \eta^2 \pi^2}{L^2}} \cos \left( \frac{2 n \pi x}{L}\right),
\eeq
where one recognizes a Jacobi theta function. The above expression makes clear that
$$
\frac{1}{L}\int_{-L/2}^{L/2} S(x) dx =1,
$$
since all the terms in the sum cancel. This applies as well to $S(y)$ and $S(z)$, so that the integral of the total source term on the fundamental cell $[-L/2,L/2]^3$ is also unity. \\
\\
The source term given by the Fourier series in  Eq.~ (\ref{SourceFourier}) represents therefore a proper regularization of the point-like lattice Universe we considered thus far. Moreover, the derivation of the solution is not affected by this change, and one can apply directly the same method as previously: the solution will remain the same, except that $f(\x)$ needs now to be replaced by $f_{\eta}(\x)$ given by
\begin{equation}
f_{\eta}(\x)=\frac{1}{\pi}\sum_{\mathbf{n}\in\Z^{3}_{*}}\frac{e^{-\frac{\pi^{2}|\mathbf{n}|^{2}\eta^{2}}{L^{2}}}}{|\mathbf{n}|^{2}}e^{\frac{2\pi}{L}i\mathbf{n}.\x}.
\end{equation}
We see that $\eta$ acts as a UV cut-off, suppressing all the terms in the series with $|\mathbf{n}|^{2}\gg\left(\frac{L}{\pi\eta}\right)^{2}$. A natural cut-off is provided by the Schwarzschild radius of the masses: $\eta=2M$, in which case we would need the first $10^{9}$ terms of each sum. Of course, since we are interested in the exterior solution, we may not need to go that far. A galaxy is approximately $10^{6}$ times bigger than its Schwarzschild radius, therefore, for an exterior solution, we may take $\eta\sim 10^{6}M$, and then, we only need the first 200 terms of each sum. 

\section{The fluid approximation}
\label{sec4}
In order to characterize the dynamics of the model we calculate the rate of expansion between two masses. Let us consider two  masses on the x-axis (all the other axes are equivalent, by symmetry), separated by a coordinate distance $NL$, for $N$ an integer. The physical distance between the two masses is given by
\begin{equation}
l(t)=\int_{0}^{NL}\sqrt{g_{xx}}dx.
\end{equation}
Then, expanding the square root to order $(M/L)^{3/2}$ we find
\begin{eqnarray}
a(t) &\equiv& \frac{l(t)}{NL}\nonumber \\
&=&1+\eps \mafrac \sqrt{\frac{G M}{L^3}} t - \frac{2 \pi G M t^2}{3 L^3} + \frac{4 \pi \eps}{9} \mafrac \left(\frac{GM}{L^3}\right)^{3/2} t^3,
\end{eqnarray}
because the part of the metric that has a spatial dependence is periodic, of period $L$, and where we introduced the effective scale factor $a(t)$ of the model. The Hubble flow defined by $H(t) = a'(t)/a(t)$ is then found to be (at order $M^{3/2}$)
\beq
H(t) = \eps \mafrac \sqrt{\frac{G M}{L^3}} - \frac{4 \pi G M t}{L^3} + 4 \sqrt{6} \pi^{3/2} \eps \left(\frac{GM}{L^3}\right)^{3/2} t^2
\eeq
Then we identify the initial expansion rate at $t=0$, 
\beq
H_0 = \eps \mafrac \sqrt{\frac{G M}{L^3}}
\eeq
where we now choose the solution $\eps=1$, ie. $H_0 >0$. This is Friedman equation for a flat Universe filled with dust with density $\rho = M/L^3$: $H_0^2= 8 \pi G M/ 3 L^3$. Replacing in the expression for $H(t)$, one finds 
\beq
\label{Hfinal}
H(t) =  H_0 -\frac{3}{2} H_0^2 t+\frac{9}{4} H_0^3 t^2 +\mathcal{O}(H_0^4)
\eeq
and this corresponds to a flat FLRW model with equation of state $w=0$. Thus, the model with discrete masses on a cubic lattice is identical to a FLRW model with dust, with the corresponding energy density. This means that one cannot distinguish the distribution of mass (localized or smeared in an homogeneous manner) from purely kinematical considerations. This is in agreement with results obtained in the Lindquist-Wheeler approximation \cite{LindWheel,Clifton:2009jw}. As mentioned in the introduction, this also agrees with the results of \cite{Clifton:2012qh} where it is found that the bigger the number of masses, the smaller the amount of backreaction produced at a given time:  here we have an infinite number of masses, and a zero backreaction. Of course, this result is only valid at order $(M/L)^{3/2}$, that is at the lowest non-linear orders; mode couplings at higher orders might alter the picture and create some backreaction. The study of these higher orders is left for future works.
A calculation of the averaged scale factor in the standard way it is done in the backreaction context \cite{Buchert:1999er}, i.e. by using volumes averages, lead to exactly the same result, provided the domain respects the symmetry of the lattice (which is necessary in the averaging context anyway, since the domain is a Lagrangian one).

\section{Conclusions}
\label{sec5}
In this paper we derived a new solution for a lattice cubic Universe that is exact up to order $M/L$ in the lattice parameter\footnote{Actually the solution extends to order $3/2$, but we expect some of the integration constants and/or free functions found at that level to be fixed by the next order of perturbation. Still, Einstein equations are solved up to order $3/2$ by Eq.~(\ref{mainresultmetric}), and, for instance, Eq.~(\ref{Hfinal}) shows that the effective Hubble constant in the lattice universe does match with its FLRW counterpart up to order $3/2$.}. The choice for the toroidal symmetry was somehow arbitrary to begin with. It is clear that the technique could be straightforwardly extended to cover more general ``crystals" -- or Bravais lattices -- without any expected new difficulties\footnote{At least for flat 3-geometry, to which we restricted ourselves through our choice for the Fourier expansions.}. This shall be explored in future works.\\
\\
In any event, the present solution Eq.~(\ref{mainresultmetric}) for the cubic lattice is a significant progress towards the understanding of the effects of inhomogeneities on the late Universe. Indeed the solution covers the whole space-time directly (or at least a significant part of it, given by $\Delta x \ll H^{-1}$, $\Delta t \ll H^{-1}$) without the need for gluing together elementary cells, thus eluding the intrinsic difficulties of the Lindquist-Wheeler scheme.\\
\\
There are two issues behind the late time inhomogeneities in the Universe: first, there is the question of their effect on the dynamics of the Universe itself. Second, it is necessary, especially in the era of precision cosmology, to quantify the effects of inhomogeneities upon the observables and in particular on the propagation of light rays; see a very interesting, recent paper on the subject \cite{Bolejko:2012ue}. In this paper we focused only on the first question and showed that the dynamics of the lattice Universe does justify the validity of the fluid approximation, in agreement with previous studies. In particular, our solution presents no backreaction at all, although it should be reminded that it is only a perturbative solution, valid for a limited amount of cosmic time, at a given order in the ratio $M/L$. Another limitation of this model is also clearly its regularity: in the real Universe, objects are structured in a more complex network made of galaxies, clusters of galaxies, filaments and large voids. Improving on this issue could be done in our model by introducing imperfections in the lattice and by studying their behaviours, as it is commonly done in solid state physics, through the study of defaults in crystal. This could be the subject of a future study.\\
\\
The question of observables in such models remains to be addressed. Is the propagation of light in the model with localized masses different from its FLRW counterpart? From the form of the metric, we should expect corrections at order $M/L$ at least, due to the presence of the anisotropic function $f$. Also, and more within an astrophysical context, what are the effects of the inhomogeneities on the local dynamics (e.g. on the rotation of spiral galaxies)? Preliminary studies indeed indicate that the function $f$ found in the solution is genuinely anisotropic, even at very short scales, i.e. around the masses themselves. Does this affect the local dynamics, and to what order of magnitude? We refer the reader to a forthcoming publication \cite{JPBJL} on these issues for the lattice Universe.

\ack
J.-P. B. is FSR/COFUND postdoctoral researcher at naXys and thanks F. Orieux for having suggested the use of the convolution theorem in Section 3. J. L. acknowledges seminal discussions with J.-P. Uzan in the early stages of this study.

\appendix
\section*{Appendix: Most general (scalar) solution up to order $M^{3/2}$.}
\setcounter{section}{1}
The following metric at order $\left(M/L\right)^{3/2}$ solves Einstein equations with source term given by Eq.~(\ref{Tab2}). We dismissed the tensorial perturbations at any order of interest, and the vectorial part is vanishing. We are thus left with scalar perturbations alone, and the metric is
\begin{eqnarray}
g_{00}=-1 \nonumber \\
g_{0i}=0 \nonumber\\
\label{EqFullMetric}
g_{ij}= \delta_{ij} \left[ 1+ 2\sqrt{\frac{GM}{Lc^2}}\left( B +  \eps \mafrac \frac{c t}{L}\right) \right. \nonumber \\
\left.+\frac{2GM}{L c^2}\left(\beta + f(\x) +\eps B\sqrt{\frac{2 \pi}{3}} \frac{c t}{L} + \frac{2 \pi c^2 t^2}{3 L^2} \right) \right.\nonumber\\
 \left.+  2 \left(\frac{GM}{Lc^2}\right)^{3/2}\left\{B f(\mathbf{x}) + \Upsilon + \eps \mafrac U(\mathbf{x}) +\frac{c t}{L}\left( \varpi + 2 \eps \mafrac  f(\mathbf{x})\right) \right. \right.\nonumber\\
\left. \left. - \frac{2 \pi B}{3} \frac{c^2 t^2}{L^2} - \frac{2 \pi \eps}{9}\mafrac \frac{c^3 t^3}{L^3}\right\} \right] \nonumber\\
+\frac{2 G M}{L c^2}\partial_{ij}^2\left(\frac{c^2 t^2}{2} f(\mathbf{x}) +L c t U(\mathbf{x}) + L^2 V(\mathbf{x})\right)\nonumber\\
+ 2\left(\frac{GM}{Lc^2}\right)^{3/2}\partial^{2}_{ij}\left[L^2 S(\x) +Lc t R(\x) + \frac{c^2 t^2}{2}\left(2 \eps\mafrac U(\x)  -B f(\x)\right)\right.\nonumber\\
\left. + \eps\sqrt{\frac{2 \pi}{3}} \frac{ c^3 t^3}{3 L} f(\x)\right].
\end{eqnarray}
The quantities $B$, $\beta$, $\Upsilon$, and $\varpi$ in Eq.~(\ref{EqFullMetric}) are constants of integration, undetermined at order $\left(M/L\right)^{3/2}$. One would need to go to higher orders to determine them. The functions $U, V, R, S$ are arbitrary periodic functions with a vanishing zero-mode.

\section*{References}
\bibliographystyle{unsrt}
\bibliography{Torus_biblio}

\begin{thebibliography}{10}

\bibitem{Clifton:2011jh}
Timothy Clifton, Pedro~G. Ferreira, Antonio Padilla, and Constantinos Skordis.
\newblock {Modified Gravity and Cosmology}.
\newblock {\em arXiv:1106.2476 [astro-ph.CO]}, 2011.

\bibitem{Celerier:1999hp}
Marie-Noelle Celerier.
\newblock {Do we really see a cosmological constant in the supernovae data?}
\newblock {\em Astron.Astrophys.}, 353:63--71, 2000.

\bibitem{Alnes:2005rw}
Havard Alnes, Morad Amarzguioui, and Oyvind Gron.
\newblock {An inhomogeneous alternative to dark energy?}
\newblock {\em Phys.Rev.}, D73:083519, 2006.

\bibitem{Enqvist:2007vb}
Kari Enqvist.
\newblock {Lemaitre-Tolman-Bondi model and accelerating expansion}.
\newblock {\em Gen.Rel.Grav.}, 40:451--466, 2008.

\bibitem{GarciaBellido:2008nz}
Juan Garcia-Bellido and Troels Haugboelle.
\newblock {Confronting Lemaitre-Tolman-Bondi models with Observational
  Cosmology}.
\newblock {\em JCAP}, 0804:003, 2008.

\bibitem{February:2009pv}
Sean February, Julien Larena, Mathew Smith, and Chris Clarkson.
\newblock {Rendering Dark Energy Void}.
\newblock {\em Mon.Not.Roy.Astron.Soc.}, 405:2231, 2010.

\bibitem{Clarkson:2010ej}
Chris Clarkson and Marco Regis.
\newblock {The Cosmic Microwave Background in an Inhomogeneous Universe - why
  void models of dark energy are only weakly constrained by the CMB}.
\newblock {\em JCAP}, 1102:013, 2011.

\bibitem{Moss:2010jx}
Adam Moss, James~P. Zibin, and Douglas Scott.
\newblock {Precision Cosmology Defeats Void Models for Acceleration}.
\newblock {\em Phys.Rev.}, D83:103515, 2011.

\bibitem{Moss:2011ze}
James~P. Zibin and Adam Moss.
\newblock {Linear kinetic Sunyaev-Zel'dovich effect and void models for
  acceleration}.
\newblock {\em Class.Quant.Grav.}, 28:164005, 2011.

\bibitem{Wiltshire:2011vy}
David~L. Wiltshire.
\newblock {What is dust? - Physical foundations of the averaging problem in
  cosmology}.
\newblock {\em Class.Quant.Grav.}, 28:164006, 2011.

\bibitem{Ell84}
G.~F.~R. {Ellis}.
\newblock {Relativistic cosmology - Its nature, aims and problems}.
\newblock In {B.~Bertotti, F.~de Felice, \& A.~Pascolini}, editor, {\em General
  Relativity and Gravitation Conference}, pages 215--288, 1984.

\bibitem{Buchert:1999er}
Thomas Buchert.
\newblock {On average properties of inhomogeneous fluids in general relativity.
  I: Dust cosmologies}.
\newblock {\em Gen. Rel. Grav.}, 32:105--125, 2000.

\bibitem{Buchert:1999mc}
Thomas Buchert.
\newblock {On average properties of inhomogeneous cosmologies}.
\newblock {\em gr-qc/0001056}, 1999.

\bibitem{Buchert:2001sa}
Thomas Buchert.
\newblock {On average properties of inhomogeneous fluids in general relativity:
  Perfect fluid cosmologies}.
\newblock {\em Gen. Rel. Grav.}, 33:1381--1405, 2001.

\bibitem{Clarkson:2011zq}
Chris Clarkson, George Ellis, Julien Larena, and Obinna Umeh.
\newblock {Does the growth of structure affect our dynamical models of the
  universe? The averaging, backreaction and fitting problems in cosmology}.
\newblock {\em Rept.Prog.Phys.}, 74:112901, 2011.

\bibitem{Clifton:2009jw}
Timothy Clifton and Pedro~G. Ferreira.
\newblock {Archipelagian Cosmology: Dynamics and Observables in a Universe with
  Discretized Matter Content}.
\newblock {\em Phys.Rev.}, D80:103503, 2009.

\bibitem{Clifton:2009bp}
Timothy Clifton and Pedro~G. Ferreira.
\newblock {Errors in Estimating $\Omega_\Lambda$ due to the Fluid
  Approximation}.
\newblock {\em JCAP}, 0910:026, 2009.

\bibitem{LindWheel}
R.~W. Linquist and J.~A. Wheeler.
\newblock {Dynamics of a Lattice Universe by the Schwarzschild-Cell Methods.}
\newblock {\em Rev. Mod. Phys}, 29:432, 1957.

\bibitem{Uzan:2010nw}
Jean-Philippe Uzan, George~F.R. Ellis, and Julien Larena.
\newblock {A two-mass expanding exact space-time solution}.
\newblock {\em Gen.Rel.Grav.}, 43:191--205, 2011.

\bibitem{Clifton:2012qh}
Timothy Clifton, Kjell Rosquist, and Reza Tavakol.
\newblock {An exact quantification of backreaction in relativistic cosmology}.
\newblock {\em arXiv:1203.6478 [gr-qc]}, 2012.

\bibitem{Yoo:2012jz}
Chul-Moon Yoo, Hiroyuki Abe, Ken-ichi Nakao, and Yohsuke Takamori.
\newblock {Black Hole Universe: Construction and Analysis of Initial Data}.
\newblock {\em arxiv:1204.2411 [gr-qc]}, 2012.

\bibitem{Korotkin:1994dw}
D.~Korotkin and H.~Nicolai.
\newblock {A Periodic analog of the Schwarzschild solution}.
\newblock {\em gr-qc/9403029}, 1994.

\bibitem{Clifton:2010fr}
Timothy Clifton.
\newblock {Cosmology Without Averaging}.
\newblock {\em Class.Quant.Grav.}, 28:164011, 2011.

\bibitem{Crandall}
R.~E. Crandall and J.~F. Delord.
\newblock {The potential within a crystal lattice}.
\newblock {\em J. Phys. A: Math. Gen.}, 20:2279--2292, 1987.

\bibitem{Crandall2}
R.~E. Crandall and J.P Buhler.
\newblock {Elementary function expansions for Madelung constants}.
\newblock {\em J. Phys. A: Math. Gen.}, 20:5497--5510, 1987.

\bibitem{Marcos:2006cn}
Bruno Marcos, T.~Baertschiger, M.~Joyce, A.~Gabrielli, and F.~Sylos~Labini.
\newblock {Linear perturbative theory of the discrete cosmological n-body
  problem}.
\newblock {\em Phys.Rev.}, D73:103507, 2006.

\bibitem{Bolejko:2012ue}
Krzysztof Bolejko and Pedro~G. Ferreira.
\newblock {Ricci focusing, shearing, and the expansion rate in an almost
  homogeneous Universe}.
\newblock {\em arxiv:1204.0909 [astro-ph.CO]}, 2012.

\bibitem{JPBJL}
J.-P. Bruneton and J.~Larena.
\newblock {Observables in a lattice Universe}.
\newblock {\em In preparation}.

\end{thebibliography}

\end{document}